\documentclass[reprint, onecolumn,superscriptaddress,amsmath,amssymb,
 aps,12pt]{revtex4-1}

\usepackage{amssymb}

\usepackage{color}
\usepackage{chemarr}
\usepackage{subfigure} 
\usepackage{graphicx}
\usepackage{epsf,mathtools}
\usepackage[outdir=./]{epstopdf}
\usepackage{epsfig}
\usepackage{dcolumn}
\usepackage{bm}
\usepackage{natbib}
\usepackage[british]{babel}

\newcommand{\mtf}{\text{mRNA}_{\text{TF}}}
\newcommand{\tf}{\text{TF}}
\newcommand{\micro}{\text{miRNA}}
\newcommand{\mrnap}{\text{mRNA}_{\text{P}}}
\newcommand{\complex}{\text{C}}
\newcommand{\protein}{\text{P}}

\newcommand{\daTF}{\text{DNA}_{a}^{TF}}
\newcommand{\diTF}{\text{DNA}_{i}^{TF}}
\newcommand{\daP}{\text{DNA}_{a}^{P}}
\newcommand{\diP}{\text{DNA}_{i}^{P}}

\newcommand{\cell}{\textrm{cell}}

\begin{document}




\title{Noise Processing by MicroRNA-Mediated Circuits: the Incoherent Feed-Forward Loop, Revisited}


\author{Silvia Grigolon}

\thanks{S. Grigolon and F. Di Patti contributed equally to this work}

\affiliation{The Francis Crick Institute, Lincoln's Inn Fields Laboratory, 44 Lincoln's Inn Fields, London WC2A 3LY, United Kingdom}
\email{silvia.grigolon@gmail.com}

\author{Francesca Di Patti}

\thanks{S. Grigolon and F. Di Patti contributed equally to this work}
\affiliation{Dipartimento di Fisica e Astronomia, Universit\`{a} degli Studi di Firenze, Sesto Fiorentino, Italy and INFN, Sezione di Firenze}
\email{f.dipatti@gmail.com}

\author{Andrea De Martino}

\thanks{A. De Martino and E. Marinari contributed equally}
\affiliation{Soft \& Living Matter Lab, Institute of Nanotechnology (CNR-NANOTEC), Consiglio Nazionale delle Ricerche, Rome, Italy; Center for Life Nano Science@Sapienza, Istituto Italiano di Tecnologia, Rome, Italy; and Human Genetics Foundation, Torino, Italy}
\email{andrea.demartino@roma1.infn.it}

\author{Enzo Marinari}

\thanks{A. De Martino and E. Marinari contributed equally}
\affiliation{Dipartimento di Fisica, Sapienza Universit\`a di Roma, Rome, Italy and INFN, Sezione di Roma}
\email{enzo.marinari@roma1.infn.it}

\begin{abstract}
The intrinsic stochasticity of gene expression is usually mitigated in higher eukaryotes by post-transcriptional regulation channels that stabilise the output layer, most notably protein levels. The discovery of small non-coding RNAs (miRNAs) in specific motifs of the genetic regulatory network has led to identifying noise buffering as the possible key function they exert in regulation. Recent \emph{in vitro} and \emph{in silico} studies have corroborated this hypothesis. It is however also known that miRNA-mediated noise reduction is hampered by transcriptional bursting in simple topologies. Here, using stochastic simulations validated by analytical calculations based on van Kampen's expansion, we revisit the noise-buffering capacity of the miRNA-mediated Incoherent Feed Forward Loop (IFFL), a small module that is widespread in the gene regulatory networks of higher eukaryotes, in order to account for the effects of intermittency in the transcriptional activity of the modulator gene. We show that bursting considerably alters the circuit's ability to control static protein noise. By comparing with other regulatory architectures, we find that direct transcriptional regulation significantly outperforms the IFFL in a broad range of kinetic parameters. This suggests that, under pulsatile inputs, static noise reduction may be less important than dynamical aspects of noise and information processing in characterising the performance of regulatory elements.
\end{abstract}

\maketitle









\section{Introduction}

Because of the inherently stochastic nature of gene expression \cite{kepler2001stochasticity,elowitz2002, kaern2005, thattai2001, raj2008}, cells dispose of a number of mechanisms to buffer the noise generated by regulatory interactions. Noise processing in eukaryotes mainly aims at preventing the amplification of fluctuations across different regulatory steps and at stabilising the output layer (proteins, RNAs, etc.), and it is normally achieved by combining a specific regulatory circuitry with some degree of  tuning of kinetic constants. The simplest non-trivial example of a noise-processing genetic circuit is perhaps the Incoherent Feed-Forward Loop (IFFL) \cite{alon2007, mangan2003, shahrezaei2008, bialek2009, bialek2010}, in which a master transcription factor (TF) activates the expression of two  molecular species, one of which inhibits the expression of the other (the target). Fluctuations in the target level are controlled by the kinetic constants that govern the system's stochastic dynamics \cite{bartel2008}, which includes molecular synthesis and  degradation steps as well as binding-mediated target repression. States for which the target level is more stable than what would be achieved in a direct regulator-target circuit lacking the intermediate repressor can generically be obtained by selecting specific ranges for kinetic rates. Very recently, this type of mechanism has been analysed in detail to clarify the role of microRNAs  (miRNAs) \cite{ambros1993, lee2001, alvarez2005, bartel2004, bartel2009} as noise-buffering agents in the post-transcriptional regulatory machinery of higher eukaryotes \cite{osella2011, ebert2012, levine2007, shimoni2007, valencia2006}. The fact that  miRNA-mediated IFFLs --where a microRNA plays the role of the repressor-- are over-represented motifs in their transcriptional regulatory network strongly suggests that static noise reduction might explain, at least in part, why this class of non-coding RNAs is so ubiquitous \cite{alon2007, tsang2007}. Theoretical {\it in silico} studies \cite{osella2011} and experimental {\it in vitro} work \cite{siciliano2013,schmiedel2015} have indeed confirmed that the miRNA-mediated IFFL can, in certain conditions, outperform more direct regulatory circuits in generating a stable protein output. 

This work aims at adding a further element to the characterisation of the noise-buffering capacity of the IFFL. We shall in particular address the question of how the latter is affected by transcriptional on/off noise at the level of the modulator (the TF). Intermittency in transcription is a well documented phenomenon \cite{raser2004control,chubb2006transcriptional,raj2006stochastic,harper2011dynamic,corrigan2014regulation} that can be driven both by intrinsic factors, like the shuttling of TFs into and out of the cell nucleus, small TF copy numbers or the peculiar thermodynamics of TF-DNA interaction \cite{gerland2002}, as well as by external signals \cite{corrigan2014regulation}. If TF-DNA binding/unbinding events are much faster than the time scales characterising RNA synthesis, it is usually safe to assume that transcription occurs continuously \cite{osella2011, bosia2012}. However, it is known that bursting severely hampers static noise buffering in simple miRNA-target modules \cite{mehta2008quantitative}. For this reason, we shall introduce and solve a model for a generalised version of the IFFL that accounts for on/off noise in transcription and that, in different limits, allows to recover the behaviour of three different regulatory circuits (a direct transcriptional regulatory module, a miRNA-mediated IFFL, and a direct module with a self-inhibiting output), whose performances we shall compare. In brief, our findings show that, in presence of intermittent transcription, the IFFL is robustly outperformed by the simpler, direct modulator-target scheme. This suggests that, as far as noise buffering is central, network architecture may not be the key to control static gene expression noise when the upstream modulator is intermittently transcribed. On the other hand, it is of crucial importance when on/off transcription noise is smoothed either due to time-scale separation or simply because typical times during which transcription is active are much longer than those over which the target level stabilises.

This scenario was mostly obtained by numerical simulations performed via the Gillespie algorithm \cite{gillespie}. For analytical validation, we resorted to  van Kampen's system-size expansion \cite{vanKampen}, a widely used approximation method for the Master Equation. Our approach differs from that employed in \cite{osella2011} and allows to study noise-buffering in slightly different conditions. While being hard to generalise beyond the Gaussian fluctuation regime, its main advantage is that it can be easily extended to cope with more complex circuits. We shall therefore also briefly outline the van Kampen's expansion solution for the standard IFFL model. 

\section{Results}

\subsection{miRNA-mediated IFFL: setup}

\begin{figure}

\begin{center}
\includegraphics[width=13cm]{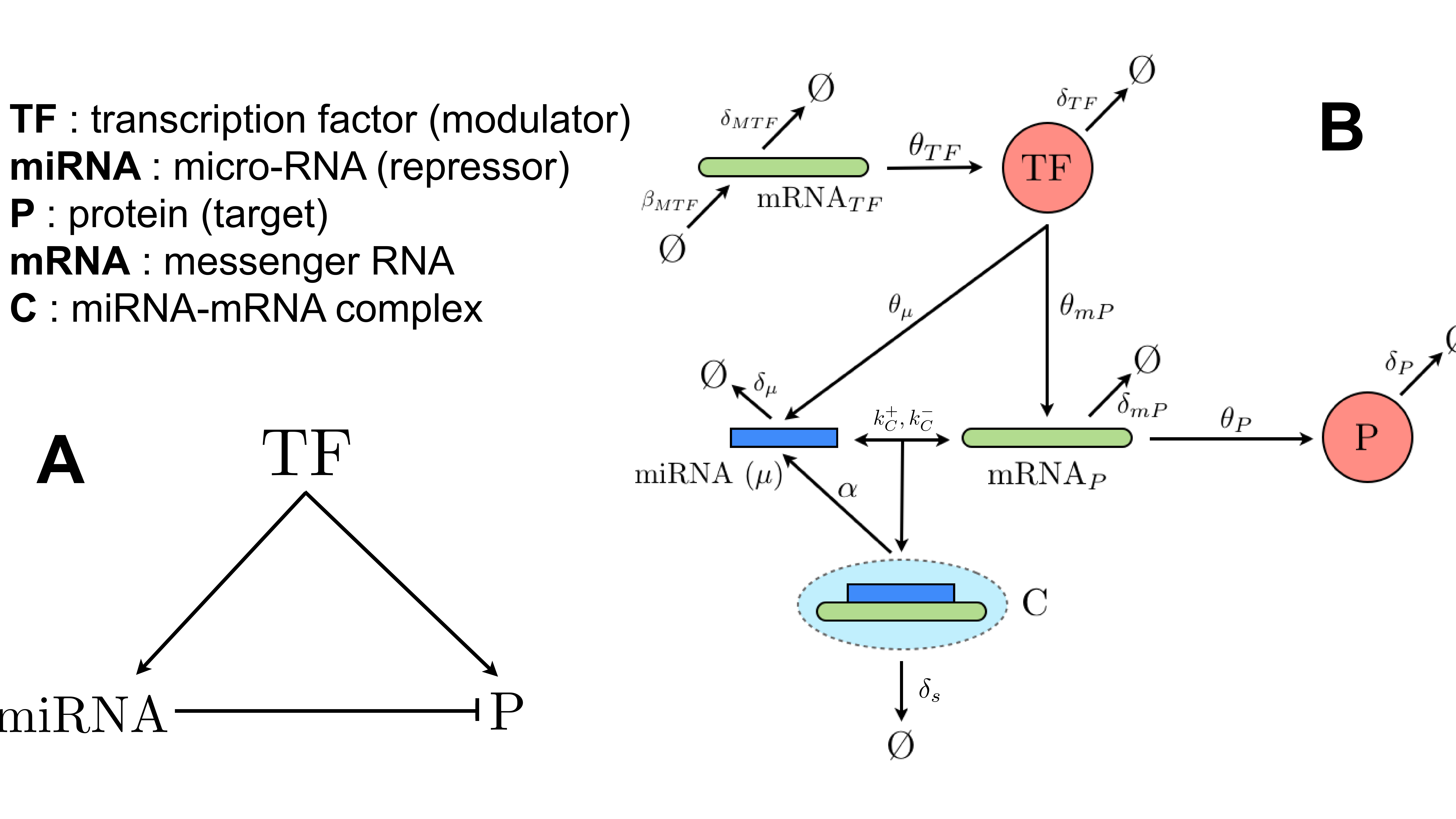}
\caption{\label{fig:schemes} \textbf{MiRNA-mediated IFFL.} (A) Schematic view. (b) Detailed view with each of the processes considered.}
\end{center}  
\end{figure}  

The miRNA-mediated IFFL is an elementary post-transcriptional regulatory circuit that accounts for the interactions between a master  TF (the modulator), a miRNA, and a target protein \cite{osella2011}. In short (see Fig. \ref{fig:schemes}A), the modulator activates the synthesis of both the miRNA and the target, which is in turn repressed by the miRNA. A detailed depiction (see Fig. \ref{fig:schemes}B) includes transcription of miRNA and messenger RNAs (mRNAs), target and modulator synthesis via translation of the mRNA substrates, complex formation and degradation, and target repression mediated by miRNA-mRNA binding that sequesters the target's mRNA thereby inhibiting translation \cite{bartel2008, bartel2009}. (In reality, miRNA-mRNA complex formation is preceded by several catalysed steps leading to the miRNA being loaded onto a specific protein complex; for simplicity, we shall ignore these steps in what follows.)  We shall denote the modulator and the target proteins, as well as the complex, by capital letters (TF, P and C, respectively), whereas we shall use the notation $\mtf$ and $\mrnap$ for the mRNAs. The processes lumped at the modulator node (with their respective rates) can be written as 
\begin{gather}
\varnothing \xrightleftharpoons[\delta _{mTF}]{\beta_{mTF}} \mtf, \label{birthMTF}  \\
\mtf               \stackrel{\theta _{TF}}{\longrightarrow}            \mtf + \tf, \label{transTF}\\
\tf                  \stackrel{\theta _{mP}}{\longrightarrow}           \mrnap + \tf,  \label{transMP}\\
\tf                 \stackrel{\theta _{\mu}}{\longrightarrow}          \micro + \tf, \label{transMU}\\
\tf                 \stackrel{\delta _{TF}}{\longrightarrow}           \varnothing \label{deathTF}~~.
\end{gather}
In brief, the modulator's mRNA ($\mtf$) is transcribed at rate $\beta_{mTF}$ and decays at rate $\delta_{mTF}$. Once transcribed, it guides the synthesis of the TF (at rate $\theta_{TF}$). The modulator, in turn, fosters the transcription of the mRNA of the target at rate $\theta_{mP}$ and of the microRNA at rate $\theta_\mu$. (For simplicity, we assume that $\mrnap$ and $\micro$ can not be transcribed from alternative loci that do not require the regulator TF.)
The target's mRNA is used as a substrate for the  synthesis of P at rate $\theta_P$:
\begin{equation}
\mrnap           \stackrel{\theta _{P}}{\longrightarrow}              \mrnap + \protein \label{transP}~~.
\end{equation}
The interaction between $\micro$ and $\mrnap$ is instead described by complex formation, dissociation, catalytic decay (with miRNA re-cycling, rate $\alpha$) and stoichiometric decay (without miRNA re-cycling, rate $\delta_s$):
\begin{gather}
\micro +\mrnap \xrightleftharpoons[k_C^- ]{k_C^+} \complex,\\
\complex           \stackrel{\alpha}{\longrightarrow}              \micro,  	\label{disintegrationC}\\
\complex            \stackrel{\delta_s}{\longrightarrow}              \varnothing~~. \label{deathC}
\end{gather}
Finally, the miRNA, the target's mRNA and the target itself are degraded respectively at rates $\delta_{\mu}$, $\delta_{mP}$ and $\delta_P$, i.e.
\begin{eqnarray}
\micro \stackrel{\delta_{\mu}}{\longrightarrow}\varnothing, \label{deathmu}~~\\
\mrnap \stackrel{\delta_{mP}}{\longrightarrow}\varnothing, \label{deathmrna}~~\\
\protein \stackrel{\delta_P}{\longrightarrow}\varnothing \label{deathP}~~.
\end{eqnarray}

\begin{figure}
\begin{center}
\includegraphics[scale=0.35]{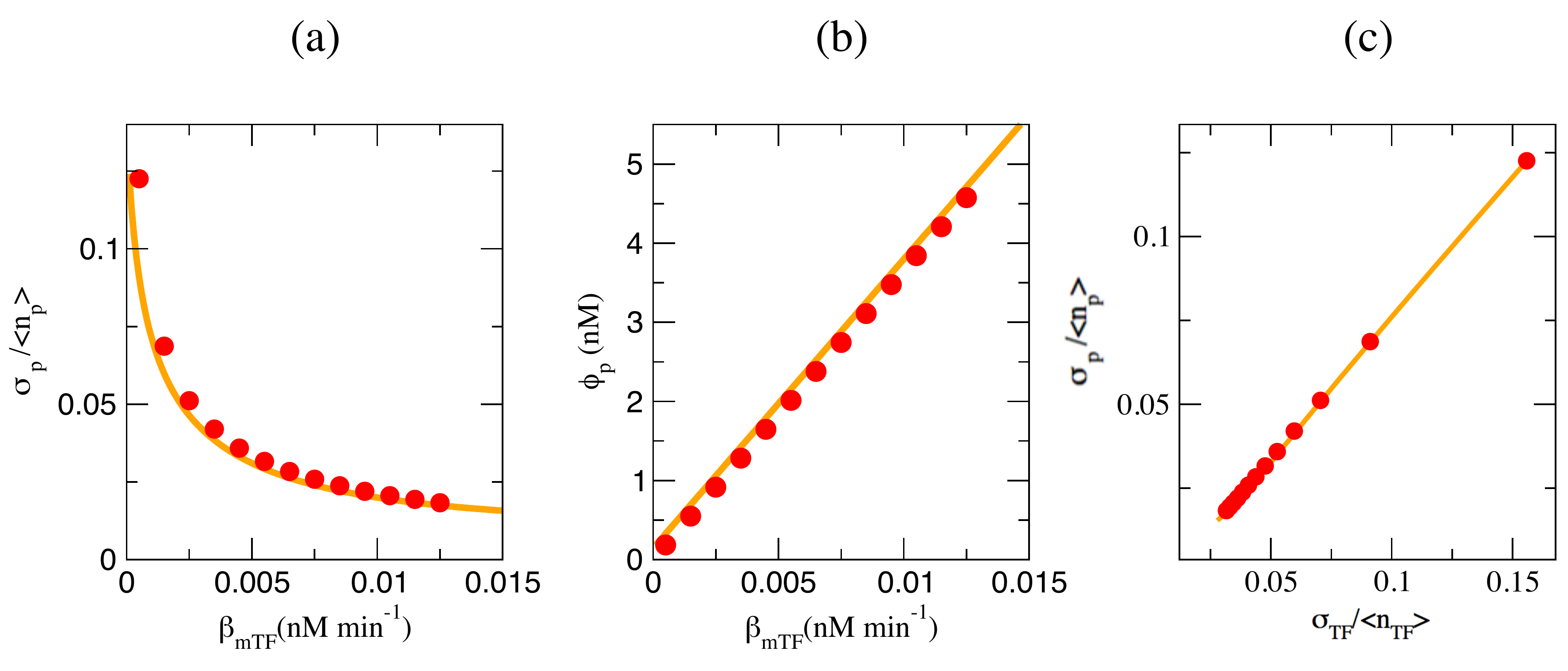} 
\caption{\textbf{Noise buffering by the miRNA-mediated IFFL (I).} Theory (lines) and simulation (markers) obtained for the IFFL. (a) Relative fluctuations in the protein number as a function of the rate constant of mRNA$_\text{TF}$ synthesis. (b)  Mean protein concentration as a function of the rate constant of mRNA$_\text{TF}$ synthesis. (c) Relative fluctuations in the protein number as a function of the relative fluctuations in the number of TFs. Averages over $10^6$ Gillespie algorithm time steps after equilibration. Rate constants are as specified in Table  \ref{tab:parameters} (see Methods), with the addition of $\theta_{TF}=0.01$ min$^{-1}$, $k_C^+/k_C^{-}=1$ $\text{nM}^{-1}$ and $\alpha=0$ min$^{-1}$}. 
\label{fig:var_vs_beta}
\end{center}
\end{figure}

By combining results from simulations performed by the Gillespie algorithm with an approximate van Kampen's expansion-based theory (see Methods) we addressed different questions, the first of which concerned the validation of our {\it in silico} results. Spanning a range of abundance for the transcription factors from roughly $50$ to around $1000$ molecules, we focused on the effects produced by the IFFL mechanism on both protein levels and fluctuations first in absence of miRNA catalytic degradation. In Fig. \ref{fig:var_vs_beta}a, we show the behaviour of the Coefficient of Variation (CV) for the target protein, i.e. $\sigma_P/\langle n_P \rangle$ with $\sigma_P$ the standard deviation and $\langle n_P\rangle$ the mean protein level, as a function of the production rate of the mRNA associated to the TF.  One sees that, as expected, relative fluctuations of the protein level decrease while its concentration $\phi_P$ increases linearly (Fig. \ref{fig:var_vs_beta}b). Likewise (Fig. \ref{fig:var_vs_beta}c), the CVs of TF and target are linearly related. In each plot, straight lines correspond to the analytical solution obtained via the van Kampen's expansion. The quality of the agreement between theory and stochastic simulations improves upon increasing the number of molecules. However, even for small values of $\phi_P$ the agreement is satisfactory. This indicates that the van Kampen's expansion, by taking the correlations between variables explicitly into account (see Methods), can lead to accurate predictions even for small numbers of molecules. On the other hand, the Gillespie algorithm results are validated through an explicit analytical approximation scheme.

\subsection{Noise buffering by the IFFL is optimised for specific miRNA levels and/or repression strengths}

\begin{figure*}
\begin{center}
\includegraphics[scale=0.4]{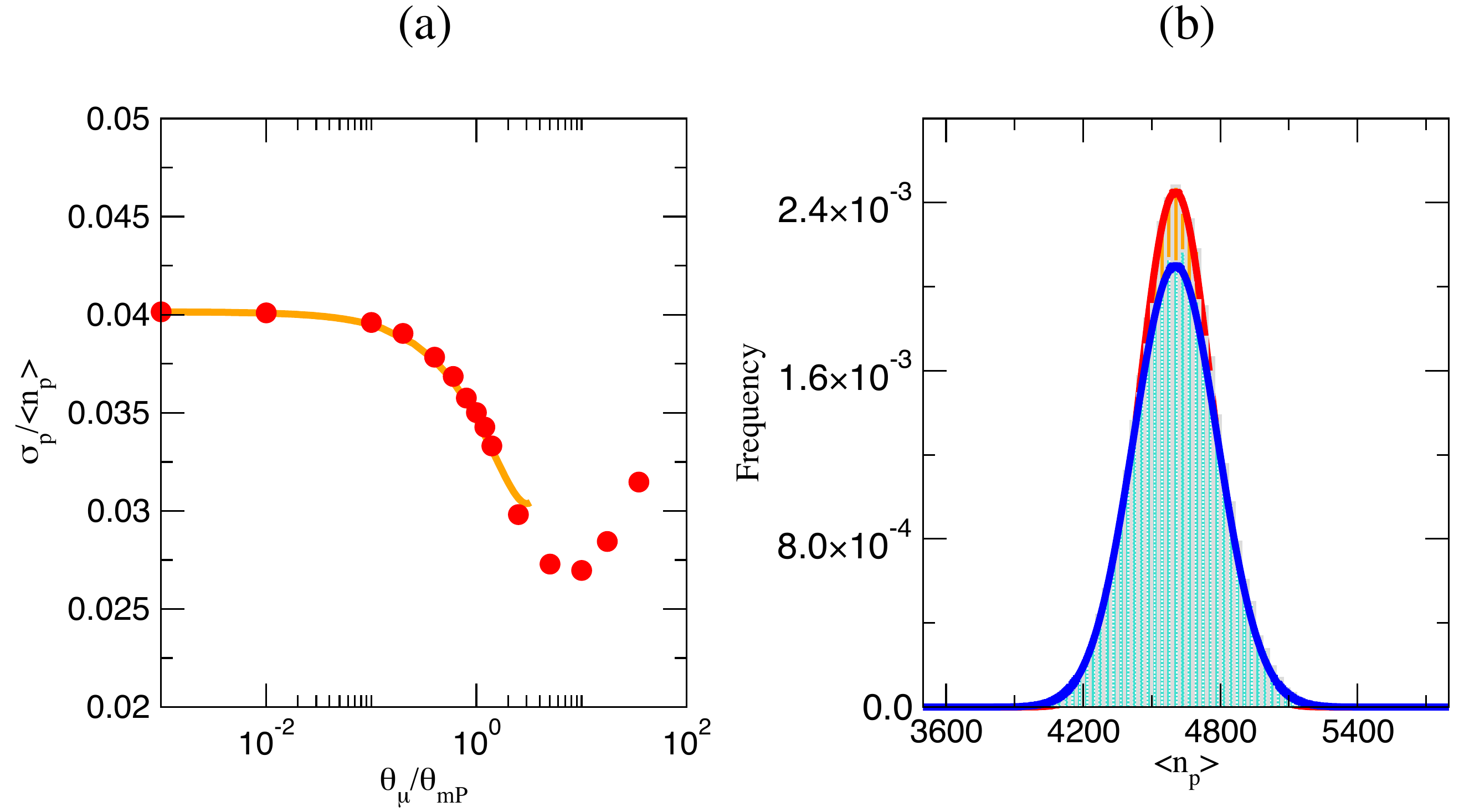}\\~\\
\includegraphics[scale=0.4]{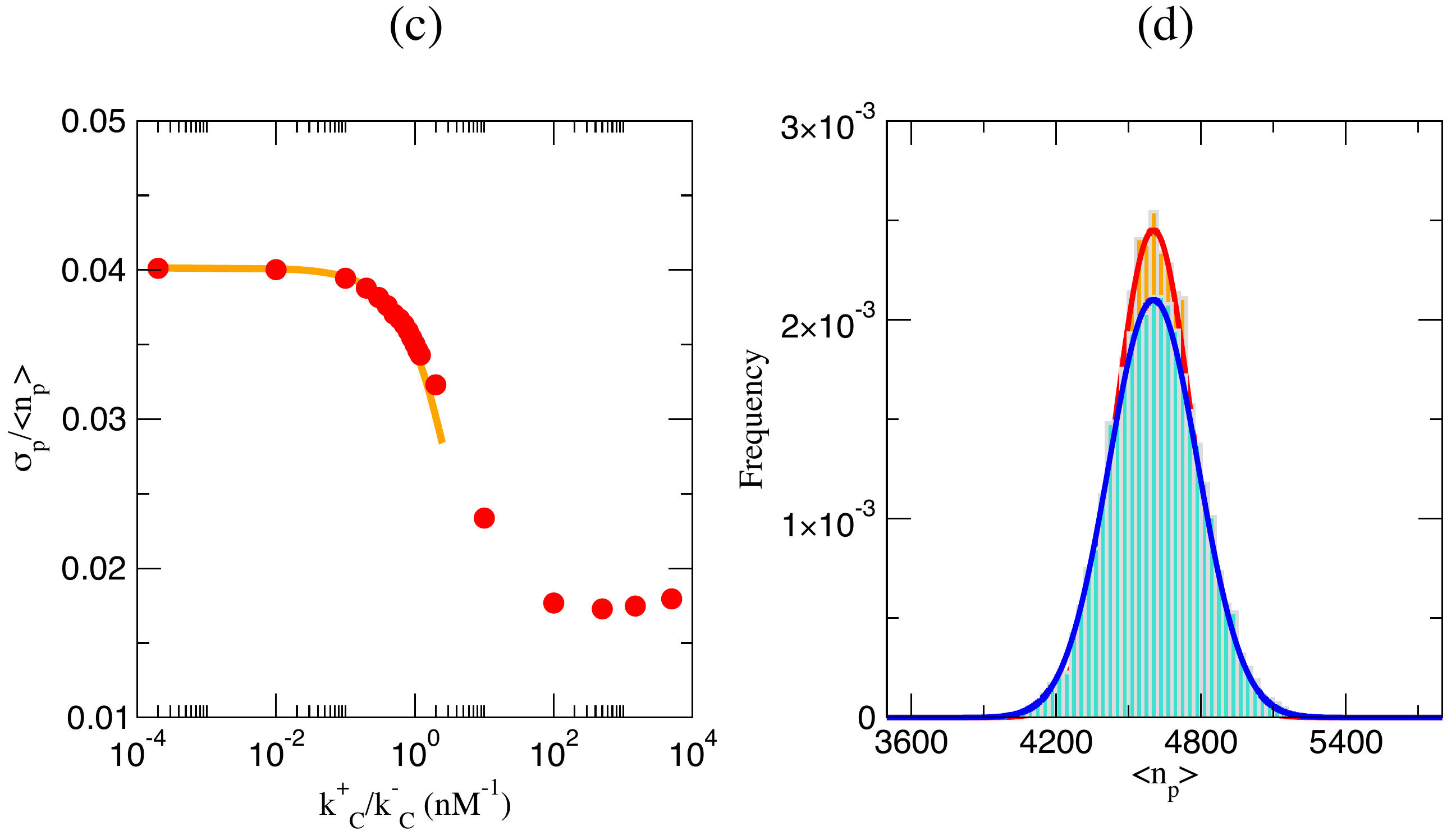}
\caption{\label{fig:var_vs_mu} \textbf{Noise buffering by the miRNA-mediated IFFL (II).} (a) Theory (lines) and simulations (markers) for the relative fluctuations generated by the IFFL in the target number upon varying the synthesis rate of miRNA ($\theta_{\mu}$, in units of $\theta_{mP}$, this last kept constant in our simulations). (b) Corresponding distributions of the target copy number obtained for fixed $\frac{\theta_{\mu}}{\theta_{mP}}=1$ (orange blocks) and $\frac{\theta_{\mu}}{\theta_{mP}}=10^{-4}$ (blue blocks). Parameters are as in Table  \ref{tab:parameters} (see Methods), with $\beta_{mTF} = 5 \cdot 10^{-3}$ nM $\cdot$ min$^{-1}$, $\frac{k^+_C}{k^-_C}=1$ nM$^{-1}$ and $\alpha=0$ min$^{-1}$. (c) Theory (lines) and simulations (markers) for the relative fluctuations generated by the IFFL in the target number upon varying the binding affinity between miRNA and the target mRNA. (d) Corresponding distributions of the target copy number obtained for fixed  $\frac{k^+_C}{k^-_C}=1$ nM$^{-1}$ (orange blocks) and $\frac{k^+_C}{k^-_C}=10^{-4}$ nM$^{-1}$ (blue blocks). Averages over $10^6$ Gillespie algorithm time steps after equilibration.  Parameters are as in Table  \ref{tab:parameters} (see Methods), with $\beta_{mTF} = 5 \cdot 10^{-3}$ nM $\cdot$ min$^{-1}$, $\theta_{\mu}=\theta_{mP}$ and $\alpha=0$ min$^{-1}$. 
}
\label{fig:var_vs_muandkc}
\end{center}
\end{figure*}

We now focus on the noise buffering capacity of the IFFL. At odds with the model considered in \cite{osella2011}, where repression is described by a Hill-like function, here we assume a simpler scenario in which concentration changes are governed by the law of mass action, so that miRNA and target can bind at rate $k^+_C$ and unbind at rate $k^{-}_C$. For simplicity, neither degradation terms for the complex are, for the moment, taken into account ($\delta_s=0$) nor catalytic decay for the target ($\alpha=0$). A regime of optimal noise buffering can be identified upon varying the miRNA transcription rate and/or the affinity between  miRNA and target at fixed target copy numbers, i.e., keeping the target transcription rate constant. In such a way, the output mean level of proteins is kept constant, allowing a consistent comparison for fluctuations. By the former route, i.e. by increasing the miRNA population in the system, protein fluctuations are found to be minimised in a specific range of values for miRNA concentration, which appears to be centred  around miRNA transcription rates roughly 10 times faster than those of the target's mRNA (see Fig. \ref{fig:var_vs_mu}a and b). 
(Notice that, when the number of miRNAs becomes too large, non-linear effects introduced by the interaction with the target become non-negligible and our van Kampen's expansion solution breaks down.) A very similar picture can be obtained by tuning the miRNA-mRNA binding rates, as reported in Figures \ref{fig:var_vs_mu}c and \ref{fig:var_vs_mu}d. In particular, higher binding rates (i.e. stronger repression) lead to smaller relative fluctuations and signatures of a minimum appear close to $k^+_C/k^-_C \simeq 10^2 \,\mbox{nM}^{-1}$. Upon increasing $k^+_C/k^-_C$ further, the Gillespie algorithm slows down considerably since miRNA-mRNA interactions become dominant. As in the previous case, the van Kampen's expansion solution can follow simulations only up to the point where non-linear effects can be neglected. 

In summary, through miRNA activity, fluctuations in the output layer can be reduced by up to 50\% compared to those characterising the input layer (obtained in absence of miRNAs, i.e., for $\theta_\mu\to 0$), in agreement with results obtained in \cite{osella2011} for a slightly different repression mechanism. While the quantitative result is parameter-dependent, this scenario is very robust at a qualitative level.

\begin{figure*}
\begin{center}
\includegraphics[scale=0.42]{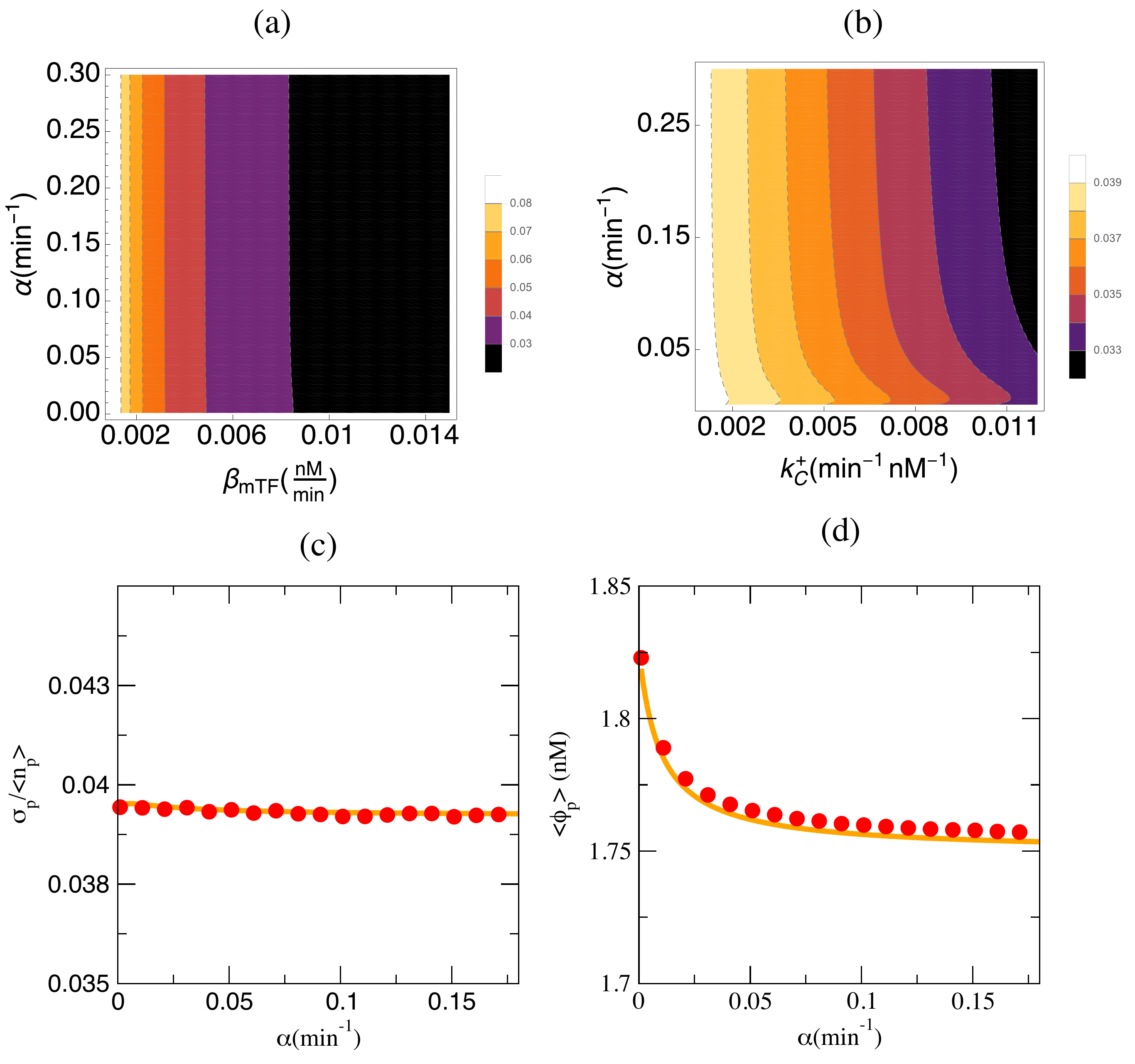}
\caption{\textbf{Effect of catalytic complex decay on the noise buffering capacity of the IFFL.} (a) Heat map of theoretical protein CVs obtained upon varying both the mRNA transcription rate $\beta_{mTF}$ and the catalytic rate $\alpha$. Kinetic parameters are as in Table \ref{tab:parameters} (see Methods), with $\frac{k^{+}_C}{k^{-}_C}=10^{-1}$ nM$^{-1}$ and $\theta_{\mu}=\theta_{mP}$. (b) Same as in (a) but now keeping $\beta_{mTF}=5 \times 10^{-3}$ $\text{nM} \cdot \text{min}^{-1}$ constant and varying the miRNA-mRNA coupling rate $k^{+}_C$. c) and d) Simulated (markers) and theoretical (line) protein CVs (c) and protein concentration (d) as a function of the catalytic decay  rate $\alpha$ of the miRNA-mRNA complex. Averages over $10^6$ GA time steps after equilibration. Parameters are as in Table  \ref{tab:parameters} (see Methods), with $\frac{k^{+}_C}{k^{-}_C}=10^{-1}$ nM$^{-1}$, $\beta_{mTF}=5 \times 10^{-3}$ nM $\cdot$ min$^{-1}$ and $\theta_{\mu}=\theta_{mP}$.}
\label{fig:heatmaps}
\end{center}
\end{figure*}

\subsection{miRNA re-cycling has a weak noise-suppressing role}

miRNAs can plausibly act  not only by sequestering the mRNA in a complex but also by catalysing its degradation, possibly without leading to its own destabilisation. This `catalytic' channel of miRNA action can be implemented in the model by simply switching on the reaction associated to the rate $\alpha$. This leads to a degradation of the target mRNA and then to a change in the average protein concentration at the steady state. miRNAs are instead fully re-cycled in the process, i.e. they re-enter the pool of free molecules after the complex is degraded. By varying the strength of catalysis, however, we only observed a weak effect on protein fluctuations, described in Fig. \ref{fig:heatmaps}, as the output's CV generically decreases as the TF transcription rate and/or the strength of repression increase. This type of effect however is hard to distinguish from the relative noise reduction due to the high numbers of molecules (Fig.\ref{fig:heatmaps}a). The competition between the catalytic channel and the pure binding-unbinding leads, in the most favourable case, to a 15\% noise reduction roughly (Fig. \ref{fig:heatmaps}b). We conclude therefore that the catalytic channel only contributes weakly to the IFFL's noise buffering capacity. 


\subsection{Generalised Feed-Forward Loop}

In order to compare the noise-buffering performance of the IFFL against other types of circuits and/or noise sources, it is convenient to introduce a generalised version of the feed-forward loop that accommodates more ingredients and by which one may recover simpler modules by turning specific reaction rates on or off. The aim is to further dissect the functionality of microRNAs under transcriptional bursting by comparing IFFL to other topologies, be they negative feedback-like or unregulated ones. To achieve this, we focus here on two additional ingredients, namely  transcriptional bursts at the level of the modulator (an extra source of stochasticity) and target self-inhibition (a possible alternative noise-buffering mechanism).

\paragraph{Transcriptional bursting}

So far, we have implicitly accounted for the noise related to the finite size of the system and the discreteness of molecules while neglecting altogether the possibility that transcription suffers from on/off noise due to the binding/unbinding of the TFs controlling modulator synthesis to the DNA promoter region \cite{elowitz2010}. Such events are usually assumed to take place on time scales much shorter than those that characterise transcription, so that for many purposes the latter can be assumed to occur at a constant rate. If the number of TFs is sufficiently high, the binding probability (which can be roughly considered to be a sigmoidal function of the TF level) is indeed essentially constant and transcription from the promoter always occurs at the largest possible rate. When one is interested in probing the system's behaviour on shorter time scales or when the population of transcription factors is not sufficiently large, promoter switching noise (under which the promoter flips between a transcribing and an idle state) should not be neglected. In order to include this mechanism to the IFFL, it suffices to replace Eqs. (\ref{birthMTF})-(\ref{transTF}) with (see also \cite{raj2006stochastic})
\begin{gather}
\daTF         \stackrel{\theta_{TF}}{\longrightarrow}               \tf + \daTF  ~~,  \label{eq:transTFnoise1}  \\
\diTF          \stackrel{k_{TF}^+}{\longrightarrow}       \daTF    ~~,\label{eq:transTFnoise2} \\
\daTF          \stackrel{k_{TF}^-}{\longrightarrow}      \diTF   ~~.\label{eq:transTFnoise3} 
\end{gather}
In short, the TF can be synthesized only if the necessary transcriptional machinery is in the active state  ($\daTF$), i.e. when all the required transcription factors are bound to the correct promoter. In turn, DNA may switch to an inactive state ($\diTF$) at rate $k_{TF}^-$. The reverse off-on transition is instead assumed to happen at rate $k_{TF}^+$. Denoting by $n_{FTa}$ and $n_{TFi}$ the number of promoters in the active and inactive state, respectively, one must additionally impose that $n_{TFa} +n_{TFi} =1$ at all times, since we assume that the TF can be transcribed from a single promoter so that $n_{FTa}\in\{0,1\}$ (and likewise for $n_{FTi}$). We shall refer to the set of chemical rules given by Eq.s (\ref{transMP})-(\ref{deathP}) and (\ref{eq:transTFnoise1})-(\ref{eq:transTFnoise3}) as the bursty-FFL  (BFFL). 
\paragraph{Target self-repression}

In addition, we want to analyse how noise can be buffered by an alternative, perhaps more intuitive mechanism. Cells must often  respond rapidly  to changing environmental conditions. The principal way through which  the cells can quickly adjust their protein levels is the enzymatic breakdown of RNA transcripts and existing protein molecules. This raises the question of whether a self-repression mechanism implemented by the target itself, through which it would inhibit its own expression, could be able to provide a tighter control of fluctuations than the BFFL.  One way by which the self-inhibition may be implemented is by the binding of the target to its own promoter, which in turn blocks accessibility to either TFs or RNA polymerase. In such a case, self-repression would become stronger as the target level increases. To analyse this scenario, one should replace Eq. (\ref{transMP}) with 
\begin{eqnarray}
\tf + \daP   & \stackrel{\theta_{mP}}{\longrightarrow}          &     \tf +\daP + \mrnap,   \\
\daP  +\protein   & \stackrel{k_{r}^+}{\longrightarrow}     &    \diP,   \\
\diP    &\stackrel{k_{r}^-}{\longrightarrow}     &    \daP +\protein,    \label{eq:selfNoise}        
\end{eqnarray}
where $\daP$ symbolises that self-repression is not active and transcription of the target's mRNAs can occur, while $\diP$ denotes an inactive state due to target self-inhibition. We shall refer to this module as the Self-Inhibiting Target (SIT). As in the previous case, the variables  $n_{Pa}$ and  $n_{Pi}$ denoting, respectively, the number of active and inactive promoters, are assumed to take the values $1$ and $0$ only, so that $n_{Pa} +n_{Pi}=1$ is an extra constraint to be enforced.

\paragraph{Description of the GFFL}

\begin{figure}
\begin{center}
\includegraphics[scale=0.62]{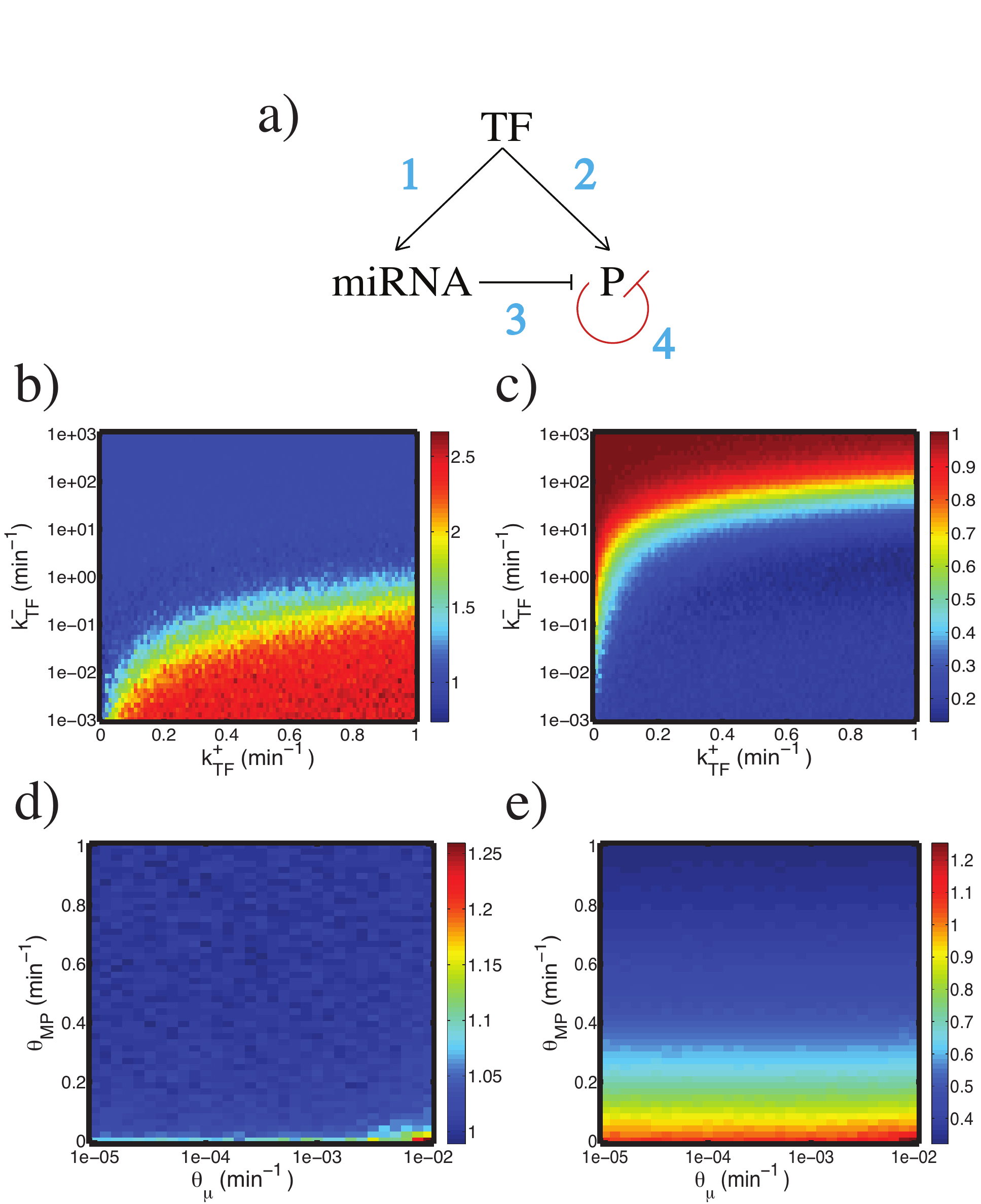} 
\caption{\textbf{Generalised FFL and transcriptional burst effect on noise buffering.} (a) Schematic representation of the generalised FFL (GFFL). The BFFL is made by the links 1,2 and 3, the SIT by the links 2 and 4 while the pure TF only by the link 2. (b)-(c) Heat map for $\text{CV}^{BFFL}/\text{CV}^{DTR}$ (b), and $\text{CV}^{BFFL}/\text{CV}^{SIT}$ (c). The chemical rates are specified in Tab. \ref{tab:parameters}, except for $\theta_{TF}=0.4$ min$^{-1}$, $k_C^+=0.01$ min$^{-1} \cdot$ nM$^{-1}$, $k_C^-=0.0001$ min$^{-1}$, $\alpha=0.001$ min$^{-1}$, $k_r^+=100$ min$^{-1} \cdot$ nM$^{-1}$ and $k_r^-=0.01$ min$^{-1}$. (d)-(e) Heat map for $\text{CV}^{BFFL}/\text{CV}^{DTR}$ (d), and $\text{CV}^{BFFL}/\text{CV}^{SIT}$ (e). The chemical rates are specified in Tab. \ref{tab:parameters}, except for $\theta_{TF}=0.4$ min$^{-1}$, $k_C^+=0.01$ min$^{-1} \cdot$ nM$^{-1}$, $k_C^-=0.0001$ min$^{-1}$, $\alpha=0.001$ min$^{-1}$, $k_r^+=100$ min$^{-1} \cdot$ nM$^{-1}$, $k_r^-=0.01$ min$^{-1}$, $k_{TF}^+=0.8$ min$^{-1}$ and $k_{TF}^-=200$ min$^{-1}$.\label{fig:GFFL}}
\end{center}  
\end{figure}

The schematics of the Generalised Feed Forward Loop (GFFL) merging the BFFL and SIT  is shown in Fig. \ref{fig:GFFL}a.   Working with the whole set of reactions (\ref{transMU})-(\ref{deathP}) and  (\ref{eq:transTFnoise1})-(\ref{eq:selfNoise}), and setting to zero some of the parameters of the full model, it is possible to describe the dynamics of the  different circuits that we wish to compare. For $k_{C}^+=k_{C}^-=\theta_{\mu}=\delta_{\mu}=0$ one gets the SIT, while the choice $k_{r}^+=k_{r}^-=0$ and $n_{Pa}\equiv 1$ leads to the BFFL. On the other hand, the straightforward case in which the protein expression is directly controlled by a TF (direct transcriptional regulation, or DTR) can be obtained by setting  $k_{r}^+=k_{r}^-=k_{C}^+=k_{C}^-=\theta_{\mu}=\delta_{\mu}=0$ and $n_{Pa} \equiv 1$.
\subsection{Direct transcriptional control outperforms the IFFL under bursty transcriptional inputs}

Because of the need to introduce the constrained binary variables $n_{TFa}$, $n_{TFi}$,  $n_{Pa}$, $n_{Pi}$ in the analysis, analytical approaches to the Master Equation through the van Kampen's expansion are in this case prevented. Our results therefore rely on stochastic simulations via Gillespie algorithm only. By analogy with the IFFL, we monitored the normalised CV of proteins  $\sigma_p/\langle n_p\rangle$ for different choices of the parameters.  In the following we will denote by CV$^{BFFL}$, CV$^{SIT}$ and   CV$^{DTR}$ the normalised CV of, respectively, the BFFL, the SIT and DRT model.  

To evaluate the role of transcriptional noise,   in Fig. \ref{fig:GFFL}  we compared the three circuits against changes in  $k_{TF}^+$ and $k_{TF}^-$, the parameters that control the magnitude of bursting activity. In order to make consistent comparisons, we checked that the mean level of proteins in each circuit stays of the same order of magnitude upon varying any parameter. Panel (b) shows the ratio $\text{CV}^{BFFL}/\text{CV}^{DTR}$. One sees that direct transcriptional modulation leads to systematically smaller CVs, although when the rate of promoter inactivation is sufficiently high and transcriptional bursts are rare the performances of the two circuits become similar. On the contrary, panel (c) compares the BFFL against the SIT and shows that miRNA-mediated control is generically more effective than self-inhibition, unless the promoter inactivation rate is much larger than the activation rate, in which case the two circuits generate similar fluctuations. 


To further investigate the regions where the three schemes appear to have the same level of effectiveness in buffering the noise, we fixed the value of the transcriptional noise to $k_{TF}^+=0.8$ min$^{-1}$ and $k_{TF}^-=200$ min$^{-1}$ (for which $\text{CV}^{BFFL}/\text{CV}^{DTR}\simeq 1$ and $\text{CV}^{BFFL}/\text{CV}^{SIT}\simeq 1$), and we varied $\theta_{mP}$ and $\theta_{\mu}$. While the ratio $\text{CV}^{BFFL}/\text{CV}^{DTR}$, displayed in Fig.\ref{fig:GFFL}d, shows very little dependence on the both the miRNA and mRNA$_\text{P}$ levels, $\text{CV}^{BFFL}/\text{CV}^{SIT}$ (Fig.\ref{fig:GFFL}e) turns out to depend strongly on  the value of  $\theta_{mP}$. Generically, for large enough $\theta_{mP}$,  the ratio is less than one, indicating more  efficient noise buffering by the BFFL with respect to the SIT. However, when the transcription rate $\theta_{mP}$ becomes sufficiently small, the trend inverts and a SIT leads to a more pronounced noise reduction.

\section{Discussion}

Stabilising the protein output is one of the central goals of the regulatory machinery of cells. In recent years, small non-coding RNAs, such as miRNAs, have been found to be involved in post-transcriptional regulation by specifically silencing mRNA translation. \emph{In silico} implementations of overrepresented genetic motifs \cite{osella2011} and \emph{in vitro} experiments on synthetic circuits \cite{siciliano2013} have elucidated the potential of miRNAs as protein noise buffering agents. The miRNA-mediated IFFL, in particular, has attracted much attention in these respects. The ability of the IFFL to buffer noise is apparent when the strength of the miRNA-target coupling is tuned in a relatively small functional range, outside of which the circuit either tends to amplify fluctuations (strong coupling limit) or produces a straightforward Poissonian statistics for the output variable (weak coupling limit). This suggests that miRNA-mediated circuits can be tuned to optimally process high-frequency noise. It is unclear whether the same performance can be achieved when transcriptional noise is involved. In this work we have indeed shown that noise buffering by the IFFL operating in the optimal regime can be severely hampered by the presence of transcriptional bursts. In particular, direct transcriptional control outperforms the IFFL when the promoter inactivation rate is sufficiently low. When the promoter inactivation rate is much larger than the rate of its inverse, finally, noise buffering appears to be network-independent as the BFFL, DTR and SIT circuits generate similar relative fluctuations.

While most transcriptional activity appears to occur continuously, intermittent bursts induced, e.g. by nucleocytoplasmatic TF transport, are increasingly being investigated especially in relation to their potential functional role. Previous work has presented evidence that pulsatile transcription can significantly limit the ability of miRNAs to control fluctuations within a simple modulator-target architecture \cite{mehta2008quantitative}. This work provides further support to this scenario by showing that noise buffering efficiency by more involved regulatory elements is tightly coupled to the strength and frequency of bursts. From a viewpoint of static noise reduction, frequent bursts would favour the selection of direct transcriptional control over other modules. In the other extreme, protein fluctuations appear to be very weakly sensitive on network architecture when the frequency of bursting is sufficiently low. 

On one hand, these results might bring further support to the idea that, while undoubtedly useful for reducing fluctuations in some case, microRNAs may carry out a more subtle --and hard to assess quantitatively-- functional role, possibly as a cross-talk intermediary between distinct targets \cite{pandolfi2011, ala2013integrated,figliuzzi2013, figliuzzi2014rna,jens2015competition,carla2015}. On the other, though, by showing that topology is not a key element in determining how well regulatory elements can dampen fluctuations in a bursty transcriptional regime, they suggest that dynamical aspects of noise processing may be more important than static noise buffering in certain conditions. Intermittent transcriptional signals can indeed transmit potentially useful information dynamically, for instance in terms of (mean) bursting frequencies, maximal transcription rates or possibly encoded in the (exponential) distributions of activation times. Our results show that downstream molecules could only exploit these specific signals at the expense of static noise reduction. While research on  dynamical information flow in the context of biochemical networks or gene regulation is in its infancy \cite{tostevin2009mutual,nemenman2012gain,mancini2013time}, it is likely that these aspects will play a major role in understanding  how and why specific regulatory input signals and circuit topologies are coupled in gene expression.

\section{Methods}

\subsection{van Kampen's expansion for the IFFL}

The mass-action kinetics of the system specified by the reactions (\ref{birthMTF})-(\ref{deathP}) can be described by the following equations for the concentrations of the various molecular species ($\phi_X=$ concentration of species $X$) involved:
\begin{gather}
	\dot{\phi}_{mTF} = \beta_{mTF}-\delta_{mTF}{\phi_{mTF}}\nonumber,\\
	\dot{\phi}_{TF} = \theta_{TF}{\phi_{mTF}}-\delta_{TF}{\phi_{TF}}\nonumber,\\
	\dot{\phi}_{mP} = \theta_{mP}{\phi_{TF}}-\delta_{mP}{\phi_{mP}}-k_C^+{\phi_{mP} \phi_{\mu}}+k_C^- {\phi_C}\nonumber,\\
	\dot{\phi}_{\mu}= \theta_{\mu} {\phi_{TF}}-\delta_{\mu}{\phi_{\mu}}-k_C^+{\phi_{mP} \phi_{\mu}}+(k_C^- + \alpha){\phi_C}, \label{eq:mfFFL} \\
	\dot{\phi}_{C} = k_C^+{\phi_{mP} \phi_{\mu}}-(k_C^- + \alpha+\delta_s){\phi_C}\nonumber,\\
	\dot{\phi}_{P} = \theta_{P} {\phi_{mTF}}-\delta_{P}{\phi_{P}}\nonumber.
\end{gather}
Here we focus for simplicity on the case $\alpha=0$ and $\delta_s=0$, where the steady state reduces to 
\begin{gather}
\phi^*_{mTF}=\frac{\beta_{mTF}}{\delta_{mTF}} ~~~~~,~~~~~
\phi^*_{TF}=\frac{\theta_{TF}}{\delta_{TF}} \phi^*_{mTF} \nonumber,\\
\phi^*_{\mu}=\frac{\theta_{\mu}}{\delta_{\mu}} \phi^*_{TF}~~~~~,~~~~~
\phi^*_{mP}=\frac{\theta_{mP}}{\delta_{mP}} \phi^*_{TF}\label{eq:epFFL},\\
\phi^*_{C}=\frac{k_C^+}{k_C^-} \phi^*_{\mu} \phi^*_{mP}~~~~~,~~~~~
\phi^*_{P}=\frac{\theta_P}{\delta_P} \phi^*_{mP}~~. \nonumber
\end{gather}
Accounting for the molecular noise induced by discreteness requires tackling the Master Equation associated to the system. Denoting by $n_X$ the number of molecules of species $X$ and by $\mathbf{n}=\{n_X\}$ the corresponding vector, with $X\in\{mTF, TF, \mu, mP, C, P\}$, the mass-action dynamics of the stochastic model is fully described by the following ME for the probability  $P(\mathbf{n},t)$ to observe the system in state $\mathbf{n}$ at time $t$ ($P\equiv P(\mathbf{n},t)$):
\begin{multline}\label{eq:meFFL}
\dot{P} =\beta_{mTF} [P(n_{mTF}-1, t) - P(\mathbf{n}, t)] 
+ \frac{\delta_{mTF}}{V_{\cell}} [(n_{mTF}+1) P(n_{mTF}+1,t) - n_{mTF} P(\mathbf{n}, t)] \\
+ \frac{\theta_{TF} n_{mTF}}{V_{\cell}} [P(n_{TF}-1,t) - P(\mathbf{n},t)] 
+ \frac{\delta_{TF}}{V_{\cell}} [(n_{TF}+1) P(n_{TF}+1,t)-n_{TF} P(\mathbf{n},t)]\\
+\frac{\theta_{\mu} n_{TF}}{V_{\cell}} [P(n_{\mu}-1,t) - P(\mathbf{n},t)]
+\frac{\delta_{\mu}}{V_{\cell}} [(n_{\mu}+1) P(n_{\mu}+1,t) - n_{\mu} P(\mathbf{n},t)]\\
+ \frac{\theta_{mP}n_{TF}}{V_{\cell}}[P(n_{mP}-1,t) - P(\mathbf{n},t)]
+\frac{\delta_{mP}}{V_{\cell}} [(n_{mP}+1)P(n_{mP}+1,t)-n_{mP}P(\mathbf{n},t)]\\
+\frac{k_C^+}{{V^2_{cell}}} [(n_{mP}+1)(n_{\mu}+1)P(n_C -1, n_{mP}+1, n_{\mu}+1, t) -n_{mP} n_{\mu} P(\mathbf{n},t)]\\
+ \frac{k_C^-}{V_{\cell}} [(n_{C}+1) P(n_C +1, n_{mP}-1, n_{\mu}-1, t)-n_C P(\mathbf{n},t)]\\
+\frac{\theta_P n_{mP}}{V_{\cell}} [P(n_P-1,t) - P(\mathbf{n},t)]
+\frac{\delta_P}{V_{\cell}} [(n_P +1) P(n_P+1,t)-n_p P(\mathbf{n},t)],
\end{multline}
where $V_{\cell}$ is the cell volume and $n_X=\phi_X \cdot V_{\cell}$.
In absence of an exact analytical solution, the main route to obtaining approximate solutions consists in focusing on the moments of $P$. In particular, one may hope to be able to compute the first (mean) and second (variance) moments of each molecular species via closed expressions. In our case, though, it can be seen that the computation of the first two moments requires knowledge of higher-order terms. 
One way to circumvent this stumbling block is to consider moments of order higher than the third to be negligible and apply van Kampen's expansion, amounting in essence to assuming that \cite{gardiner,vanKampen}
\begin{equation}\label{ansatx}
n_X=V_{\cell}\phi_X(t)+ \xi_X\sqrt{V_{\cell}}
\end{equation}
for $X\in\{mTF, TF, \mu, mP, C, P\}$. The parameter $V_{\cell}$ represents the ``system size'', i.e., in practice, the cell volume, which is assumed to be sufficiently large. In other terms, one splits the discrete variables $n_X$ into two components: a deterministic term $\phi_X$ and a fluctuating term proportional to the random variable $\xi_X$. At the leading order in $V_{\cell}$, the distributions appearing in the Master Equation are Dirac $\delta$-distributions centred around $\phi_X$. In this case, the Master Equation leads to the deterministic differential equations for the concentrations $\phi_X$ given in (\ref{eq:mfFFL}). The next-to-leading order instead corresponds to the assumption that the distributions of molecule numbers are Gaussian, centred around $\phi_X$ and with a finite variance. Since averages are fixed by the macroscopic terms, the Master Equation reduces, at this order, to an equation for the distribution of the variances, $\Pi(\boldsymbol{\xi},t)$ (with $\boldsymbol{\xi}=\{\xi_X\}$).
In terms of the van Kampen creation and destruction operators \cite{vanKampen}
\begin{equation}
\epsilon^{\pm}_X \simeq \mathbb{I} \pm V_{\cell}^{-\frac{1}{2}} \frac{\partial}{\partial \xi_X} + \frac{1}{2} V_{\cell}^{-1} \frac{\partial^2}{\partial \xi_X^2}~~,
\end{equation}
one has
\begin{equation}
P(n_X \pm 1, t) = \epsilon^{\pm}_X P(n_X,t)~~,
\end{equation}
where we are hiding the dependence on the species whose numbers are not changing. Moreover, since by inverting (\ref{ansatx}) one gets
\begin{equation}
\xi_X(n_X)=V_{\cell}^{-\frac{1}{2}}n_X+V_{\cell}^{-1}\phi_X~~,
\end{equation}
we see that distributions of molecular populations can be expressed in terms of distributions of the noise variables, i.e.
\begin{equation}
P(n_X\pm 1, t)=\Pi(\xi_X \pm V_{\cell}^{-\frac{1}{2}},t)~~.
\end{equation}
Expanding this expression for large $V_{\cell}$, we get
\begin{multline}
\Pi(\xi_X \pm V_{\cell}^{-\frac{1}{2}},t) \equiv \epsilon^{\pm}_X \Pi(\xi_X, t) \simeq \\
\simeq \Pi(\xi_X, t) \pm V_{\cell}^{-\frac{1}{2}} \frac{\partial \Pi(\xi_X, t)}{\partial \xi_X} + \frac{1}{2} V_{\cell}^{-1} \frac{\partial^2 \Pi(\xi_X, t)}{\partial \xi_X^2} + \mathcal{O}(V_{\cell}^{-\frac{3}{2}})~~.
\end{multline}
Likewise, for the left-hand-side of the Master Equation we have
\begin{equation}
\frac{dP(n_X, t)}{dt}=\frac{\partial \Pi(\xi_X, t)}{\partial t}- V_{\cell}^{\frac{1}{2}}\sum_X \dot{\phi}_X \frac{\partial \Pi(\xi_X, t)}{\partial \xi_X}~~.
\end{equation}
Re-scaling the time with $V_{\cell}$ as $t \rightarrow t/V_{\cell}\equiv\tau$, we get a Fokker-Plank equation for the fluctuations $\xi_X$, which, in the case of the IFFL, reads
\begin{equation}
\begin{split}
\frac{\partial \Pi}{\partial \tau} & = \frac{\beta_{mTF}}{2} \frac{\partial^2 \Pi}{\partial \xi_{mTF}^2} + \frac{\delta_{mTF}}{2} \phi_{mTF} \frac{\partial^2 \Pi}{\partial \xi_{mTF}^2} + \delta_{mTF} \frac{\partial}{\partial \xi_{mTF}} (\xi_{mTF} \Pi) \\
&+ \theta_{TF} \frac{\phi_{mTF}}{2} \frac{\partial^2 \Pi}{\partial \xi_{TF}^2} - \theta_{TF} \xi_{mTF} \frac{\partial \Pi}{\partial \xi_{TF}} + \delta_{TF} \frac{\partial}{\partial \xi_{TF}} (\xi_{TF} \Pi) + \frac{\delta_{TF}}{2} \phi_{TF} \frac{\partial^2 \Pi}{\partial \xi_{TF}^2} \\
& + \theta_{\mu} \frac{\phi_{TF}}{2} \frac{\partial^2 \Pi}{\partial \xi_{\mu}^2} - \theta_{\mu} \xi_{TF} \frac{\partial \Pi}{\partial \xi_{ \mu}} + \delta_{\mu} \frac{\partial}{\partial \xi_{\mu}} (\xi_{\mu} \Pi) + \frac{\delta_{\mu}}{2} \phi_{\mu} \frac{\partial^2 \Pi}{\partial \xi_{\mu}^2} \\
&+ \theta_{mP} \frac{\phi_{TF}}{2} \frac{\partial^2 \Pi}{\partial \xi_{mP}^2} - \theta_{mP} \xi_{TF} \frac{\partial \Pi}{\partial \xi_{mP}} + \delta_{mP} \frac{\partial}{\partial \xi_{mP}} (\xi_{mP} \Pi) + \frac{\delta_{mP}}{2} \phi_{mP} \frac{\partial^2 \Pi}{\partial \xi_{mP}^2}\\
& - k^+_C (\xi_{mP} \phi_{\mu} + \phi_{mP} \xi_{\mu}) \frac{\partial \Pi}{\partial \xi_C} + k^+_C \xi_{mP} \phi_{\mu} \frac{\partial \Pi}{\partial \xi_{\mu}} \\
&+ k^+_C \phi_{mP} \frac{\partial}{\partial \xi_{\mu}} (\xi_{\mu} \Pi) - k^+_C \phi_{\mu} \phi_{mP} \frac{\partial^2}{\partial \xi_C \partial \xi_{\mu}} \Pi + k^+_C \phi_{mP} \xi_{\mu} \frac{\partial \Pi}{\partial \xi_{mP}} \\
&+ k^+_C \phi_{\mu} \frac{\partial }{\partial \xi_{mP}} (\xi_{mP} \Pi) - \phi_{\mu} \phi_{mP} k^+_C \frac{\partial^2 \Pi}{\partial \xi_C \partial \xi_{mP}} \\
&+ k^+_C \phi_{\mu} \phi_{mP} \frac{\partial^2 \Pi}{\partial \xi_{\mu} \partial \xi_{mP}} + \frac{k^+_C}{2} \phi_{\mu} \phi_{mP} \frac{\partial^2 \Pi}{\partial \xi_{mP}^2}+ \frac{k^+_C}{2} \phi_{\mu} \phi_{mP} \frac{\partial^2 \Pi}{\partial \xi_{\mu}^2}+ \frac{k^+_C}{2} \phi_{\mu} \phi_{mP} \frac{\partial^2 \Pi}{\partial \xi_{C}^2}\\
& - k^-_C \xi_C \frac{\partial \Pi}{\partial \xi_{mP}} - k^-_C \phi_C \frac{\partial^2 \Pi}{\partial \xi_C \partial \xi_{mP}} - k^-_C \xi_C \frac{\partial \Pi}{\partial \xi_{\mu}} + \frac{k^-_C \phi_C}{2} \frac{\partial^2 \Pi}{\partial \xi_{\mu}^2} + \frac{k^-_C \phi_C}{2} \frac{\partial^2 \Pi}{\partial \xi_{mP}^2} \\
&+ \frac{k^-_C \phi_C}{2} \frac{\partial^2 \Pi}{\partial \xi_C^2} + k^-_C \frac{\partial}{\partial \xi_C} (\xi_C \Pi) - k^-_C \phi_C \frac{\partial^2 \Pi}{\partial \xi_C \partial \xi_{\mu}}\\
& - k^-_C \phi_C \frac{\partial^2 \Pi}{\partial \xi_{\mu} \partial \xi_{mP}} + \frac{\theta_P \phi_{mP}}{2} \frac{\partial^2 \Pi}{\partial \xi_P^2}\\
& - \theta_P \xi_{mP} \frac{\partial \Pi}{\partial \xi_P} + \delta_P \frac{\partial}{\partial \xi_P} (\xi_P \Pi) + \frac{\delta_P \phi_P}{2} \frac{\partial^2 \Pi}{\partial \xi_P^2}~~.
\end{split}
\end{equation}
Once its coefficients have been evaluated at the steady state described by (\ref{eq:epFFL}), the above equation can be used to identify a system of differential equations for the second moments of the fluctuations $\langle \xi_X \xi_Y\rangle$ (which we do not report for simplicity). Because at stationarity $\langle \xi_X \rangle = 0$~ $\forall X$, one has 
\begin{equation}
\sigma^2_X = \langle n^2_X \rangle - \langle n_X \rangle^2 = V_{\cell} \langle \xi^2_X \rangle~~,
\end{equation}
which relates the variance of $n_X$ to that of $\xi_X$. Notice that, within this approach, concentrations $\phi_X$ directly translate to number of molecules $n_X$. For instance, if $\phi_X$ is expressed in nanomolars, and $V_{\cell}$ 
is the cell volume measured in $\mu$m$^3$, one has
\begin{equation}
n_X =N_A \,V_{\cell} \,\phi_X \, 10^{-9}~ \text{molecules}~~,
\label{eq:conversion}
\end{equation} 
where $N_A$ is the Avogadro number. 

\subsection{The Gillespie algorithm}


The Gillespie algorithm \cite{gillespie} is an exact method to  simulate reaction kinetics lin well-mixed systems (though it can be generalised to non-Markovian and non well-mixed systems as well \cite{boguna2014}). It is based on a Monte Carlo procedure to generate 
the Markov process described by the Master Equation. Hereby, we shall describe how we applied it to our specific case (see \cite{gillespie} for further insights). 

In the present study, the system has been initialised in each case at the steady state (which can easily be computed from deterministic equations). 
At each time step, we have checked for physical consistency for  the number of molecules for each species (which should stay positive or zero) as well as for the the total number of molecules in the system (which, by the finiteness of the volume should not exceed a large threshold, which we set to be $10^6$, so as to guarantee that each species involved in the circuit can be comparable with experimentally estimated ones). 
After running the Gillespie algorithm for a long enough time to accumulate many reaction events, we have extracted the statistics (mean and variance mainly) for the number of molecules of each of the species. 

Parameters that are not varied across this study were obtained from experimental literature and set as in Table \ref{tab:parameters} in order to guarantee the same concentration values for the different species as in the cited experiments. Variable parameters have been changed across ranges that guarantee biologically plausible steady state concentrations. 

\begin{table}
\begin{center}
\begin{tabular}{|c|c|c|}
\hline
Parameter & Value & Description \\
\hline
$\delta_{mP}$ & $0.02 \text{ min}^{-1}$ & Target degradation rate\\ 
$\theta_{mP}$    & $0.08 \text{ min}^{-1}$ & Target expression rate\\        
$\delta_{\mu} $ & $0.015 \text{ min}^{-1}$ & miRNA degradation rate\\
$\theta_{\mu}$ &  $\theta_{mP}$  & miRNA expression rate\\
$\delta_{TF}$     & $0.01 \text{ min}^{-1}$  & master Transcription Factor degradation rate   \\
$\delta_P$ & $\delta_{TF}$   & Protein degradation rate        \\   
$\theta_P$ & $0.03 \text{ min}^{-1}$ & Protein expression rate \\
$\delta_{mTF}$ & $ \delta_{mP}$  & master Transcription Factor's mRNA degradation rate \\
$k^{-}_C$ & $0.01 \text{ min}^{-1}$ & miRNA-mRNA target dissociation constant\\
$R_{\cell}$ & $10 \mu$m & Cell radius\\
$N$ & $10^6$ molecules & Maximal number of molecules \\
  \hline
\end{tabular}
\end{center}
\caption{Values of fixed model parameters, from \cite{siciliano2013} and \cite{bionumbers}. For simplicity, the cell has been approximated as a sphere of radius $R_{\cell}$, so that $V_{\cell}=4\pi R_{\cell}^3/3$.) 
} 
\label{tab:parameters}
\end{table}

\section*{Declarations}
\section*{Author contribution statement}
Silvia Grigolon, Francesca Di Patti: Performed the experiments; Analyzed and interpreted the data; Contributed reagents, materials, analysis tools or data; Wrote the paper.
Andrea De Martino, Enzo Marinari: Conceived and designed the experiments; Analyzed and interpreted the data; Contributed reagents, materials, analysis tools or data; Wrote the paper.
Silvia Grigolon and Francesca Di Patti contributed equally to this study.
Andrea De Martino and Enzo Marinari contributed equally to this study.

\section*{Competing interest statement}
The authors declare no conflict of interest.

\section*{Funding statement}
This work was supported by the Marie Curie Training Network NETADIS (FP7, grant 290038), by Ente Cassa di Risparmio di Firenze and program PRIN 2012 funded by the Ministero dell'Istruzione, dell'Universit\`a e della Ricerca (MIUR) and by the Francis Crick Institute which receives its core funding from Cancer Research UK, the UK Medical Research Council, and the Wellcome Trust.

\section*{Additional information}
Data associated with this study has been deposited at http://chimera.roma1.infn.it/SYSBIO/.

\section*{Ackowledgements}

We thank Olivier C. Martin for useful suggestions and for a critical reading of the manuscript, as well as Carla Bosia, Matteo Figliuzzi and Andrea Pagnani for discussions.

%
%
%
%
\section*{References}
\bibliographystyle{unsrtnat}
\bibliography{bibliography}

\begin{thebibliography}{47}
\providecommand{\natexlab}[1]{#1}
\providecommand{\url}[1]{\texttt{#1}}
\expandafter\ifx\csname urlstyle\endcsname\relax
  \providecommand{\doi}[1]{doi: #1}\else
  \providecommand{\doi}{doi: \begingroup \urlstyle{rm}\Url}\fi

\bibitem[Kepler and Elston(2001)]{kepler2001stochasticity}
T.B. Kepler and T.C. Elston.
\newblock Stochasticity in transcriptional regulation: origins , consequences ,
  and mathematical representations.
\newblock \emph{Biophysical Journal}, 81\penalty0 (6):\penalty0 3116--36, 2001.

\bibitem[Elowitz et~al.(2002)Elowitz, Levine, Siggia, and Swain]{elowitz2002}
M.~B. Elowitz, A.~J. Levine, E.~D. Siggia, and P.~S. Swain.
\newblock Stochastic gene expression in a single cell.
\newblock \emph{Science}, 297\penalty0 (5584):\penalty0 1183--86, 2002.

\bibitem[Kaern et~al.(2005)Kaern, Elston, Blake, and Collins]{kaern2005}
M.~Kaern, T.C. Elston, W.J. Blake, and J.J. Collins.
\newblock Stochasticity in gene expression: from theories to phenotypes.
\newblock \emph{Nat Rev Genet}, 6:\penalty0 451--64, 2005.

\bibitem[Thattai and van Oudenaarden(2001)]{thattai2001}
M.~Thattai and A.~van Oudenaarden.
\newblock Intrinsic noise in gene regulatory networks.
\newblock \emph{PNAS}, 98\penalty0 (15):\penalty0 8614--19, 2001.

\bibitem[A.Raj and Oudenaarden(2008)]{raj2008}
A.Raj and A.van Oudenaarden.
\newblock Nature , nurture , or chance: stochastic gene expression and its
  consequences.
\newblock \emph{Cell}, 135\penalty0 (2):\penalty0 216--26, 2008.

\bibitem[Alon(2007)]{alon2007}
U.~Alon.
\newblock \emph{An introduction to systems biology}.
\newblock Chapman \& Hall/CRC, 2007.

\bibitem[Mangan and Alon(2003)]{mangan2003}
S.~Mangan and U.~Alon.
\newblock Structure and function of the feed-forward loop network motif.
\newblock \emph{PNAS}, 100\penalty0 (21):\penalty0 11980--85, 2003.

\bibitem[Shahrezaei et~al.(2008)Shahrezaei, Ollivier, and
  Swain]{shahrezaei2008}
V.~Shahrezaei, J.~F. Ollivier, and P.~S. Swain.
\newblock Colored extrinsic fluctuations and stochastic gene expression.
\newblock \emph{Mol. Syst. Biol.}, 4\penalty0 (196), 2008.

\bibitem[Walczak et~al.(2009)Walczak, Tka\u{c}ik, and Bialek]{bialek2009}
A.~M. Walczak, G.~Tka\u{c}ik, and W.~Bialek.
\newblock Optimizing information flow in small genetic networks.
\newblock \emph{Physical Review E}, 80\penalty0 (3 Pt 1):\penalty0 031920,
  2009.

\bibitem[Walczak et~al.(2010)Walczak, Tka\u{c}ik, and Bialek]{bialek2010}
A.~M. Walczak, G.~Tka\u{c}ik, and W.~Bialek.
\newblock Optimizing information flow in small genetic networks ii: Feed
  forward interactions.
\newblock \emph{Physical Review E}, 81:\penalty0 041905, 2010.

\bibitem[Baek et~al.(2008)Baek, Vill\'en, Shin, Camargo, Gygi, and
  Bartel]{bartel2008}
D.~Baek, J.~Vill\'en, C.~Shin, F.D. Camargo, S.P. Gygi, and D.P. Bartel.
\newblock The impact of microrna on protein output.
\newblock \emph{Nature}, 455\penalty0 (7209):\penalty0 64--71, 2008.

\bibitem[Lee et~al.(1993)Lee, Feinbaum, and Ambros]{ambros1993}
R.~C. Lee, R.~L. Feinbaum, and V.~Ambros.
\newblock The c. elegans heterochronic gene lin-4 encodes small rnas with
  antisense complementarity to lin-14.
\newblock \emph{Cell}, 75\penalty0 (5):\penalty0 843--54, 1993.

\bibitem[Lee and Ambros(2001)]{lee2001}
R.~C. Lee and V.~Ambros.
\newblock An extensive class of small rnas in caenorhabditis elegans.
\newblock \emph{Science}, 294\penalty0 (5543):\penalty0 862--4, 2001.

\bibitem[Alvarez-Garcia and Miska(2005)]{alvarez2005}
I.~Alvarez-Garcia and E.~A. Miska.
\newblock Microrna functions in animal development and human disease.
\newblock \emph{Development}, 132\penalty0 (21):\penalty0 4653--62, 2005.

\bibitem[Bartel(2004)]{bartel2004}
D.~P. Bartel.
\newblock Micrornas: Genomics , biogenesis , mechanism , and function.
\newblock \emph{Cell}, 116\penalty0 (2):\penalty0 281--97, 2004.

\bibitem[Bartel(2009)]{bartel2009}
D.~P. Bartel.
\newblock Micrornas: Target recognition and regulatory functions.
\newblock \emph{Cell}, 136\penalty0 (2):\penalty0 215--33, 2009.

\bibitem[Osella et~al.(2011)Osella, Bosia, Cor\`{a}, and Caselle]{osella2011}
M.~Osella, C.~Bosia, D.~Cor\`{a}, and M.~Caselle.
\newblock The role of incoherent micro{RNA}-mediated feedforward loops in noise
  buffering.
\newblock \emph{PLoS Computational Biology}, 7\penalty0 (3):\penalty0 e1001101,
  2011.

\bibitem[Ebert and Sharp(2012)]{ebert2012}
M.S. Ebert and P.A. Sharp.
\newblock Roles for micrornas in conferring robustness to biological processes.
\newblock \emph{Cell}, 149\penalty0 (3):\penalty0 515--24, 2012.

\bibitem[Levine et~al.(2007)Levine, Zhang, Kuhlman, and Hwa]{levine2007}
E.~Levine, Z.~Zhang, T.~Kuhlman, and T.~Hwa.
\newblock Quantitative characteristics of gene regulation by small {RNA}.
\newblock \emph{PLoS Biology}, 5\penalty0 (9):\penalty0 e229, 2007.

\bibitem[Shimoni et~al.(2007)Shimoni, Friedlander, Hetzroni, Niv, Altuvia,
  Biham, and Margalit]{shimoni2007}
Y.~Shimoni, G.~Friedlander, G.~Hetzroni, G.~Niv, S.~Altuvia, O.~Biham, and
  H.~Margalit.
\newblock Regulation of gene expression by small non-coding {RNA}s: a
  quantitative view.
\newblock \emph{Molecular Systems Biology}, 3:\penalty0 138, 2007.

\bibitem[Valencia-Sanchez et~al.(2006)Valencia-Sanchez, J.Liu, G.J.Hannon, and
  R.Parker]{valencia2006}
M.A. Valencia-Sanchez, J.Liu, G.J.Hannon, and R.Parker.
\newblock Control of translation and mrna degradation by mirnas and sirnas.
\newblock \emph{Genes Dev}, 20\penalty0 (5):\penalty0 515--24, 2006.

\bibitem[Tsang et~al.(2007)Tsang, Zhu, and van Oudenaarden]{tsang2007}
J.~Tsang, J.~Zhu, and A.~van Oudenaarden.
\newblock Microrna-mediated feedback and feedforward loops are recurrent
  network motifs in mammals.
\newblock \emph{Mol Cell}, 26\penalty0 (5):\penalty0 753--67, 2007.

\bibitem[Siciliano et~al.(2013)Siciliano, Garzilli, Fracassi, Criscuolo,
  Ventre, and di~Bernardo]{siciliano2013}
V.~Siciliano, I.~Garzilli, C.~Fracassi, S.~Criscuolo, S.~Ventre, and
  D.~di~Bernardo.
\newblock mirnas confer phenotypic robustness to gene networks by suppressing
  biological noise.
\newblock \emph{Nat. Commun.}, 4:\penalty0 2364, 2013.

\bibitem[Schmiedel et~al.(2015)Schmiedel, Klemm, Zheng, Sahay, Bl\"uthgen,
  Marks, and van Oudenaarden]{schmiedel2015}
J.~M. Schmiedel, S.~L. Klemm, Y.~Zheng, A.~Sahay, N.~Bl\"uthgen, D.~S. Marks,
  and A.~van Oudenaarden.
\newblock Microrna control of protein expression noise.
\newblock \emph{Science}, 348\penalty0 (6230):\penalty0 128--32, 2015.

\bibitem[Raser and O'Shea(2004)]{raser2004control}
J.~M. Raser and E.~K. O'Shea.
\newblock Control of stochasticity in eukaryotic gene expression.
\newblock \emph{Science}, 304\penalty0 (5678):\penalty0 1811--4, 2004.

\bibitem[Chubb et~al.(2006)Chubb, Trcek, Shenoy, and
  Singer]{chubb2006transcriptional}
J.~R. Chubb, T.~Trcek, S.~M. Shenoy, and R.~H. Singer.
\newblock Transcriptional pulsing of a developmental gene.
\newblock \emph{Current Biology}, 16\penalty0 (10):\penalty0 1018--25, 2006.

\bibitem[Raj et~al.(2006)Raj, C.S.Peskin, D.Tranchina, D.Y.Vargas, and
  S.Tyagi]{raj2006stochastic}
A.~Raj, C.S.Peskin, D.Tranchina, D.Y.Vargas, and S.Tyagi.
\newblock Stochastic mrna synthesis in mammalian cells.
\newblock \emph{PLoS Biol}, 4\penalty0 (10):\penalty0 e309, 2006.

\bibitem[Harper et~al.(2011)Harper, Finkenst{\"a}dt, Woodcock, Friedrichsen,
  Semprini, Ashall, Spiller, Mullins, Rand, Davis, et~al.]{harper2011dynamic}
C.~V. Harper, B.~Finkenst{\"a}dt, D.~Woodcock, S.~Friedrichsen, S.~Semprini,
  L.~Ashall, D.~Spiller, J.~J. Mullins, D.~Rand, J.~R. Davis, et~al.
\newblock Dynamic analysis of stochastic transcription cycles.
\newblock \emph{PLoS Biology}, 9\penalty0 (4):\penalty0 e1000607, 2011.

\bibitem[Corrigan and Chubb(2014)]{corrigan2014regulation}
A.~M. Corrigan and J.~R. Chubb.
\newblock Regulation of transcriptional bursting by a naturally oscillating
  signal.
\newblock \emph{Current Biology}, 24\penalty0 (2):\penalty0 205--11, 2014.

\bibitem[Gerland et~al.(2002)Gerland, Moroz, and Hwa]{gerland2002}
U.~Gerland, J.~D. Moroz, and T.~Hwa.
\newblock Physical constraints and functional characteristics of transcription
  factor - dna interaction.
\newblock \emph{PNAS}, 99\penalty0 (19):\penalty0 12015--20, 2002.

\bibitem[Bosia et~al.(2012)Bosia, Osella, Baroudi, Cor\`{a}, and
  Caselle]{bosia2012}
C.~Bosia, M.~Osella, M.~El Baroudi, D.~Cor\`{a}, and M.~Caselle.
\newblock Gene autoregulation via intronic micrornas and its functions.
\newblock \emph{BMC Systems Biology}, 6:\penalty0 131, 2012.

\bibitem[Mehta et~al.(2008)Mehta, Goyal, and Wingreen]{mehta2008quantitative}
P.~Mehta, S.~Goyal, and N.~S. Wingreen.
\newblock A quantitative comparison of srna-based and protein-based gene
  regulation.
\newblock \emph{Molecular Systems Biology}, 4:\penalty0 221, 2008.

\bibitem[Gillespie(1976)]{gillespie}
D.~T. Gillespie.
\newblock A general method for numerically simulating the stochastic time
  evolution of coupled chemical reactions.
\newblock \emph{J. Comp. Phys.}, 22:\penalty0 403--434, 1976.

\bibitem[{van Kampen}(2007)]{vanKampen}
N.~G. {van Kampen}.
\newblock \emph{Stochastic processes in Physics and Chemistry}.
\newblock North Holland, Amsterdam, 3rd edition, 2007.

\bibitem[Eldar and Elowitz(2010)]{elowitz2010}
A.~Eldar and M.B. Elowitz.
\newblock Functional roles for noise in genetic circuits.
\newblock \emph{Nature}, 467:\penalty0 167--173, 2010.

\bibitem[Salmena et~al.(2011)Salmena, Poliseno, Tay, Kats, and
  Pandolfi]{pandolfi2011}
L.~Salmena, L.~Poliseno, Y.~Tay, L.~Kats, and P.~P. Pandolfi.
\newblock A cerna hypothesis: the rosetta stone of a hidden rna language?
\newblock \emph{Cell}, 146\penalty0 (3):\penalty0 353--58, 2011.

\bibitem[Ala et~al.(2013)Ala, Karreth, Bosia, Pagnani, Taulli, L{\'e}opold,
  Tay, Provero, Zecchina, and Pandolfi]{ala2013integrated}
U.~Ala, F.~Karreth, C.~Bosia, A.~Pagnani, R.~Taulli, V.~L{\'e}opold, Y.~Tay,
  P.~Provero, R.~Zecchina, and P.~P. Pandolfi.
\newblock Integrated transcriptional and competitive endogenous rna networks
  are cross-regulated in permissive molecular environments.
\newblock \emph{PNAS}, 110\penalty0 (18):\penalty0 7154--9, 2013.

\bibitem[Figliuzzi et~al.(2013)Figliuzzi, Marinari, and Martino]{figliuzzi2013}
M.~Figliuzzi, E.~Marinari, and A.~De Martino.
\newblock Micrornas as a selective channel of communication between competing
  rnas: a steady-state theory.
\newblock \emph{Biophysical Journal}, 104\penalty0 (5):\penalty0 1203--13,
  2013.

\bibitem[Figliuzzi et~al.(2014)Figliuzzi, Martino, and
  Marinari]{figliuzzi2014rna}
M.~Figliuzzi, A.~De Martino, and E.~Marinari.
\newblock Rna-based regulation: dynamics and response to perturbations of
  competing rnas.
\newblock \emph{Biophysical Journal}, 107\penalty0 (4):\penalty0 1011--22,
  2014.

\bibitem[Jens and Rajewsky(2015)]{jens2015competition}
M.~Jens and N.~Rajewsky.
\newblock Competition between target sites of regulators shapes
  post-transcriptional gene regulation.
\newblock \emph{Nature Reviews Genetics}, 16\penalty0 (2):\penalty0 113--26,
  2015.

\bibitem[Bosia et~al.(2015)Bosia, Sgr{\`o}, Conti, Baldassi, Cavallo, Cunto,
  Turco, Pagnani, and Zecchina]{carla2015}
C.~Bosia, F.~Sgr{\`o}, L.~Conti, C.~Baldassi, F.~Cavallo, F.~Di Cunto,
  E.~Turco, A.~Pagnani, and Riccardo Zecchina.
\newblock Quantitative study of crossregulation, noise and synchronization
  between microrna targets in single cells.
\newblock \emph{arxiv.org/abs/1503.06696}, 2015.

\bibitem[Tostevin and Wolde(2009)]{tostevin2009mutual}
F.~Tostevin and P.~R.~Ten Wolde.
\newblock Mutual information between input and output trajectories of
  biochemical networks.
\newblock \emph{Physical Review Letters}, 102\penalty0 (21):\penalty0 218101,
  2009.

\bibitem[Nemenman(2012)]{nemenman2012gain}
I.~Nemenman.
\newblock Gain control in molecular information processing: lessons from
  neuroscience.
\newblock \emph{Physical Biology}, 9\penalty0 (2):\penalty0 026003, 2012.

\bibitem[Mancini et~al.(2013)Mancini, Wiggins, Marsili, and
  Walczak]{mancini2013time}
F.~Mancini, C.~H. Wiggins, M.~Marsili, and A.~M. Walczak.
\newblock Time-dependent information transmission in a model regulatory
  circuit.
\newblock \emph{Physical Review E}, 88\penalty0 (2):\penalty0 022708, 2013.

\bibitem[Gardiner(2004)]{gardiner}
C.~W. Gardiner.
\newblock \emph{Handbook of Stochastic Methods: for Physics, Chemistry and the
  Natural Sciences}.
\newblock Springer, 3rd edition, 2004.

\bibitem[Bogu$\tilde{n}$$\acute{a}$ et~al.(2014)Bogu$\tilde{n}$$\acute{a}$,
  Lafuerza, Toral, and $\acute{A}$ngeles Serrano]{boguna2014}
M.~Bogu$\tilde{n}$$\acute{a}$, L.~F. Lafuerza, R.~Toral, and
  M.~$\acute{A}$ngeles Serrano.
\newblock Simulating non-markovian stochastic processes.
\newblock \emph{Phys. Rev. E}, \penalty0 (90):\penalty0 042108, 2014.

\bibitem[Milo et~al.(2010)Milo, Jorgensen, Moran, Weber, and
  Springer]{bionumbers}
R.~Milo, P.~Jorgensen, U.~Moran, G.~Weber, and M.~Springer.
\newblock Bionumbers---the database of key numbers in molecular and cell
  biology.
\newblock \emph{Nucl. Acids Res.}, 38\penalty0 (suppl 1):\penalty0 D750--D753,
  2010.

\end{thebibliography}




\end{document}